\documentclass[twocolumn,aps,prl,superscriptaddress,10pt]{revtex4-2}
\usepackage{graphicx}
\usepackage{amsmath}
\usepackage{amssymb}
\usepackage{braket}
\usepackage{bm}
\usepackage{dcolumn}
\usepackage{mathrsfs}
\usepackage{graphicx}
\usepackage[colorlinks=true,linkcolor=blue,citecolor=blue,urlcolor=blue]{hyperref}
\usepackage{soul}
\usepackage{changes}
\usepackage{amsthm,amsmath,amssymb}
\usepackage{float}
\usepackage{mathrsfs}
\usepackage[toc,page,title,titletoc,header]{appendix}
\usepackage[percent]{overpic}  % 在导言区
\usepackage{picture}                    
\makeatother
\begin{document}

\title{A General Strategy for Realizing Mpemba Effects in Open Quantum Systems}  

\author{Yaru Liu}

\affiliation{Shenzhen Institute for Quantum Science and Engineering, Southern University of Science and Technology,
Shenzhen 518055, China}
\affiliation{Department of Physics, Southern University of Science and Technology, Shenzhen 518055, China}
\author{Yucheng Wang}
\email{wangyc3@sustech.edu.cn}
%\selectlanguage{english}%
\affiliation{Shenzhen Institute for Quantum Science and Engineering, Southern University of Science and Technology,
Shenzhen 518055, China}
\affiliation{International Quantum Academy, Shenzhen 518048, China}
\affiliation{Guangdong Provincial Key Laboratory of Quantum Science and Engineering, Southern University of Science and Technology, Shenzhen 518055, China}

%\date{\today}
\begin{abstract}
The Mpemba effect, where a state farther from equilibrium relaxes faster than one closer to it, is a striking phenomenon in both classical and quantum systems. In open quantum systems, however, the quantum Mpemba effect (QME) typically occurs only for specifically chosen initial states, which limits its universality. Here we present a general and experimentally feasible strategy to realize both QME and anti-QME. By applying a temporary bond-dissipation quench, we selectively suppresses or enhances slow relaxation modes, thereby reshaping relaxation pathways independently of both the system and the initial state. We demonstrate this mechanism in systems with dephasing and boundary dissipation, and outline feasible cold-atom implementations. Our results establish controllable dissipation as a versatile tool for quantum control, accelerated relaxation, and efficient nonequilibrium protocols.
%Our results establish controllable dissipation as a general strategy for tailoring relaxation pathways and open new possibilities for quantum control, accelerated state preparation, and thermodynamically efficient protocols.
\end{abstract}
\maketitle

\textit{Introduction.—}
Nonequilibrium physics gives rise to a variety of nontrivial and counterintuitive phenomena that continue to attract growing interest~\cite{RMP1,RMP2,RMP3,RMP4}. A striking example is the Mpemba effect~\cite{Mpemba1969,Mpemba2017,Mpemba2019,Mpemba2025,review1,review2,review3}, where hot water can freeze faster than cold water under identical conditions. While originally observed in classical systems, its quantum counterpart, the quantum Mpemba effect (QME), has recently received significant attention~\cite{review2,review3,QMEclose1,QMEclose2,QMEclose3,QMEclose4,QMEclose5,QMEclose6,QMEclose7,QMEopen1,QMEopen2,QMEopen3,QMEopen4,QMEopen5,QMEopen6,QMEopen7,QMEopen8,QMEopen8s,QMEopen9,QMEtwo1,QMEtwo2}. In isolated quantum systems, QME describes situations where symmetry is restored more quickly from a highly asymmetric initial state than from a more symmetric one, under a symmetric Hamiltonian quench~\cite{review2,review3,QMEclose1,QMEclose2,QMEclose3,QMEclose4,QMEclose5,QMEclose6,QMEclose7}. In open quantum systems, QME arises when a state initially farther from the steady state relaxes more rapidly than one closer to it, due to dissipative coupling to an environment~\cite{review2,review3,QMEopen1,QMEopen2,QMEopen3,QMEopen4,QMEopen5,QMEopen6,QMEopen7,QMEopen8,QMEopen8s,QMEopen9}.
Related anomalous relaxation phenomena also include the inverse Mpemba effect (IME), in which a colder state can heat up faster than a warmer one under a thermal quench~\cite{Mpemba2017,Mpemba2019,IME1,IME2}. These developments not only deepen our understanding of relaxation dynamics but also offer opportunities for quantum control, accelerated state preparation, and thermodynamically efficient protocols in open quantum technologies. However, because the occurrence of QME in open systems typically relies on specifically chosen initial states, its universality remains limited.

Recent experimental advances have established highly controllable platforms for probing the dynamics of open quantum systems, including cold atoms, trapped ions, photonic lattices, and superconducting circuits. In these platforms, different types of dissipation can be engineered and precisely tuned. Dissipation is no longer regarded merely as a detrimental source of decoherence but has instead emerged as a versatile resource for controlling quantum phases and driving nonequilibrium transitions~\cite{PTopen1,PTopen2,PTopen3,PTopen4,PTopen5,PTopen6,PTopen7,PTopen8,PTopen9,PTopen10,PTopen11,PTopen12,PTopen13}. This progress opens the door to investigating how carefully designed dissipative processes can fundamentally reshape relaxation pathways and uncover new regimes of quantum dynamics.

In this work, we focus on the role of controllable bond dissipation~\cite{PTopen13,PZollerB0,PZollerB1,PZollerB2,PZollerB3,PZollerB4,Marcos2012,Yusipov1,ChenS} in open quantum systems. By introducing bond dissipation during a finite time interval, we demonstrate that the relaxation pathway can be significantly altered. Specifically, suppressing slow modes accelerates relaxation of states farther from equilibrium, robustly inducing QME. Conversely, by tuning the phase parameter of the bond dissipation, slow modes can be deliberately enhanced, giving rise to what we term the anti-QME. We note that this effect is distinct from the IME: whereas IME concerns heating processes under thermal quenches, the anti-QME refers to the controlled slowing of relaxation in dissipative open quantum dynamics. Crucially, whether slow modes are suppressed or enhanced does not depend on the initial state, implying that both QME and anti-QME can be realized in a controllable manner across a broad class of systems. Our findings therefore establish a general and experimentally feasible strategy for tailoring relaxation dynamics in open quantum systems.

\textit{Controlling slow modes via bond dissipation.—}
The dynamics of a quantum system coupled to a Markovian environment 
can be described by the Lindblad master equation~\cite{GLindblad,HPBreuer},
\begin{equation}\label{eq:Lindblad}
	\frac{d\rho}{dt} = \mathcal{L}_0[\rho] 
	= -i[H,\rho] 
	+ \sum_j \Big( O^{(0)}_j \rho O^{(0)\dagger}_j  
	- \tfrac{1}{2}\{O^{(0)\dagger}_j O^{(0)}_j, \rho\}\Big),
\end{equation}
%\begin{equation}\label{eq:Lindblad}
%	\frac{d\rho}{dt} = \mathcal{L}[\rho] = -i[H,\rho] 
%	+ \sum_k \gamma_k \Big( L_k \rho L_k^\dagger 
%	- \tfrac{1}{2}\{L_k^\dagger L_k, \rho\}\Big),
%\end{equation}
where $H$ is the system Hamiltonian and $O^{(0)}_j$ are the jump operators. We focus on two common types of $O^{(0)}_j$: dephasing dissipation and boundary dissipation.
$\mathcal{L}_0$ denotes the Liouvillian in the absence of bond dissipation, with right and left 
eigenmodes $\{r_j, l_j\}$ and eigenvalues $\lambda_j$: 
$\mathcal{L}_0[r_j] = \lambda_j r_j$, $\mathcal{L}_0^\dagger[l_j] = \lambda_j^* l_j$,
normalized according to $\mathrm{Tr}[\,l_i^\dagger r_j\,] = \delta_{ij}$.
The time evolution of the density matrix can 
then be expressed as
\begin{equation}
	\rho(t) = \sum_j e^{\lambda_j t} \alpha_j r_j, \qquad 
	\alpha_j = \mathrm{Tr}[l_j^\dagger \rho(0)],
\end{equation}
which explicitly contains the steady state 
$\rho_{\mathrm{ss}} = r_0$ with $\lambda_0 = 0$. The mode with eigenvalue $\lambda_1$, having the smallest nonzero real part, usually sets the longest relaxation time, so the late-time dynamics is dominated by this slowest mode, unless its weight is negligible, in which case the next-slowest mode takes over.

\begin{figure}[!htp]
	\centering
	\includegraphics[width=0.5\textwidth]{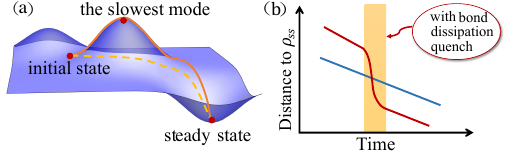}
	\caption{ Schematic illustration of the QME.  
		(a) The blue surface schematically represents the relaxation landscape, where valleys correspond to slow relaxation channels dominated by the slowest Liouvillian mode. The natural relaxation trajectory (orange solid line) follows a valley and is therefore slow, while an alternative trajectory (orange dashed line) bypasses the valley and reaches the steady state faster.
		(b) Time evolution of two initial states: although the red state starts farther from the steady state, applying bond dissipation (yellow-shaded region) accelerates its relaxation so that it overtakes the blue state, demonstrating the QME.
	}
	\label{fig1}
\end{figure}

If the weight of the initial state on this slowest mode can be reduced and redistributed into faster-decaying modes, the overall relaxation process is accelerated [Fig. \ref{fig1}(a)]. 
To achieve such control, we introduce a tunable form of bond dissipation. 
The corresponding jump operator $O^{(1)}_j$ acts on a pair of sites $j$ and $j+p$, 
with a uniform dissipation rate $\Gamma$, and is defined as~\cite{PTopen13,PZollerB0,PZollerB1,PZollerB2,PZollerB3,PZollerB4,Marcos2012,Yusipov1,ChenS}
\begin{equation}
	O^{(1)}_j = \sqrt{\Gamma}(c_j^\dagger + a\,c_{j+p}^\dagger)(c_j - a\,c_{j+p}), 
	\qquad a=\pm 1, \ p\ge 1,
	\label{eq_oj}
\end{equation}
%where $c_j$ annihilates a particle on site $j$. 
where $c_j$ ($c^\dagger_j$) annihilates (creates) a particle on lattice site $j$. This dissipation conserves the total particle number but modifies relative phases between sites separated by distance $p$. 
Depending on the parameter $a$, it drives particles into either in-phase 
($a=1$) or out-of-phase ($a=-1$) states. Such dissipative mechanisms have been proposed in cold-atom platforms~\cite{PTopen13,PZollerB0,PZollerB1,PZollerB2,PZollerB3,PZollerB4}, superconducting resonator arrays~\cite{Marcos2012}, and superconducting quantum circuits~\cite{ChenS}.

In our protocol, bond dissipation is applied only within a finite time interval $t_1<t<t_2$. During this time interval, the system involves both $O^{(0)}_j$ and $O^{(1)}_j$, with the corresponding Liouvillian denoted as $\mathcal{L}_1$.
Accordingly, the system evolves as
\begin{equation}\label{evolution}
	\rho(t) =
	\begin{cases}
		e^{\mathcal{L}_0 t}\,\rho(0), & t < t_1, \\[6pt]
		e^{\mathcal{L}_1 (t-t_1)}\, e^{\mathcal{L}_0 t_1}\,\rho(0), & t_1 < t < t_2, \\[6pt]
		e^{\mathcal{L}_0 (t-t_2)}\, e^{\mathcal{L}_1 (t_2-t_1)}\, e^{\mathcal{L}_0 t_1}\,\rho(0), & t > t_2.
	\end{cases}
\end{equation}
After the quench, the system continues to relax toward the steady state of the original Liouvillian $\mathcal{L}_0$. 
%The crucial point is that the temporary bond dissipation modifies the overlap of the evolving state with the slowest mode [Fig. \ref{fig1}(a)], thereby enabling either suppression (QME) or enhancement (anti-QME) of slow relaxation.

We next examine how a temporary bond-dissipation quench alters the original slow mode of
$\mathcal{L}_0$. Its overlap with the evolving state is characterized by
\begin{equation}
	\mu_1(t) = \mathrm{Tr}[\, l_1^\dagger \rho(t) \,].
\end{equation} 
After an initial evolution up to time $t_1$, the state is
$\rho(t_1) = e^{\mathcal{L}_0 t_1}\rho(0)$.
For a short quench of duration $\tau$, expanding to first order gives $\rho(t_1+\tau) \approx \rho(t_1) 
	+ \tau ( \mathcal{L}_1 - \mathcal{L}_0 ) \rho(t_1)$.
Projecting onto the slow mode yields
$\mu_1(t_1+\tau) \approx \mathrm{Tr}[\,l_1^\dagger \rho(t_1)\,] 
	+ \tau\,\mathrm{Tr}[\,l_1^\dagger ( \mathcal{L}_1-\mathcal{L}_0 ) \rho(t_1) \,]$.
Therefore, the change of the slow-mode amplitude during the quench is
	$\Delta \mu_1 = \mu_1(t_1+\tau)-\mu_1(t_1) 
	\approx \tau \sum_j e^{\lambda_jt_1}\alpha_j \,\mathrm{Tr}[\,l_1^\dagger ( \mathcal{L}_1-\mathcal{L}_0 ) [r_j] \,]$. This shows that the slow-mode amplitude receives contributions from all modes of $\mathcal{L}_0$, weighted by their occupations $e^{\lambda_jt_1}\alpha_j$ and the Liouville-space transfer matrix elements
$\mathrm{Tr}[\,l_1^\dagger (\mathcal{L}_1-\mathcal{L}_0)[r_j]\,]$, which quantify the coupling between mode $j$ and the original slow mode. 
As a result, weight can be transferred into the slow mode ($\Delta\mu_1>0$), leading to a longer relaxation time and the anti-Mpemba effect, or transferred out of it ($\Delta\mu_1<0$), leading to accelerated relaxation and the QME [Fig. \ref{fig1}(b)].

We now turn to two concrete examples. We emphasize that the bond dissipation is applied only for a finite time interval, and the system ultimately relaxes to the steady state of the original (unquenched) system. If the bond dissipation is maintained for too long, the system remains near the steady state of the bond-dissipation Liouvillian for an extended period, which generally differs from the steady state of the original dissipative dynamics under consideration. In this case, the bond dissipation no longer accelerates relaxation but instead prolongs the relaxation time. Therefore, we apply the bond dissipation only during a short transient interval. Unless otherwise specified, we return to open boundary conditions in the following sections.

\begin{figure}[tb]
	\centering
    \includegraphics[width=8.57cm]{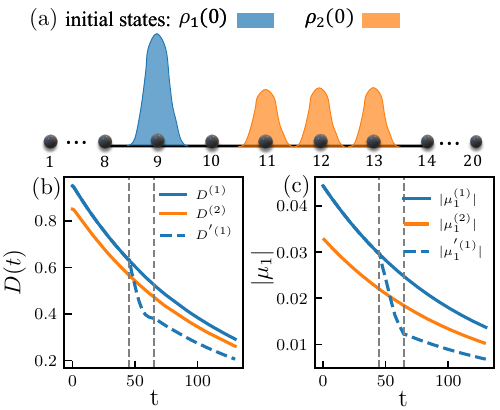}
    \caption{ QME induced by a temporary bond-dissipation quench under dephasing dissipation. (a) Initial states: $\rho_1$ (localized, far from equilibrium) and $\rho_2$ (less localized, closer to equilibrium).
    	(b) Time evolution of the trace distance $D^{(i)}(t)$, where the blue (orange) solid line corresponds to the initial state $\rho_1(0)$ [$\rho_2(0)$]. A short bond dissipation quench ($t_1=45$, $t_2=65$) accelerates the relaxation of $\rho_1$ (blue dashed line).   
    	(c) Time evolution of the slowest-mode amplitude $|\mu_1|$ corresponding 
    	to the three cases in (b). Other parameters: $\gamma^d=0.01$, $\Gamma=0.01$, $p=1$, $a=1$.
    }
	\label{fig2}
\end{figure}

\textit{Example I: Dephasing dissipation.—} As a first example, we consider uniform dephasing with rate $\gamma^d$, described by the local jump operators
\begin{equation}
O^{(0)}_j = \sqrt{\gamma^d} c^\dagger_j c_j,
\end{equation}
 which suppresses coherence on each site and drives the system into a unique steady state.
Bond dissipation $O^{(1)}_j$ is then applied only during a finite time 
interval $t_1 < t < t_2$. We fix the system size to $L=20$ and prepare two distinct 
initial states [Fig. \ref{fig2}(a)]: a fully localized state 
$\rho_1(0) = |9\rangle\langle9|$, and a three-site uniformly distributed state 
$\rho_2(0) = \tfrac{1}{3}(|11\rangle\langle11| + |12\rangle\langle12| + |13\rangle\langle13|)$. 
The time evolution $\rho(t)$ is governed by Eq.(\ref{evolution}).  

The relaxation dynamics is monitored through the trace distance~\cite{review2}
\begin{equation}
	D^{(i)}(t) = \tfrac{1}{2}\,\mathrm{Tr}\,|\rho_i(t) - \rho_{\mathrm{ss}}|,
\end{equation}
which measures the distance between the evolving state $\rho_i(t)$ and the steady 
state $\rho_{\mathrm{ss}}$, with $i=1,2$ labeling the two different initial 
conditions. We denote the trace distance and slowest-mode amplitude without bond dissipation as \( D^{(i)} \) and $|\mu_1|$, and their counterparts with a temporary bond-dissipation quench as \( D^{\prime(i)} \) and $|\mu^{\prime}_1|$. Under pure dephasing, the steady state is uniform, so $\rho_2(0)$ remains closer to equilibrium ($D^{(2)} < D^{(1)}$) throughout the evolution [Fig. \ref{fig2}(b)], and no QME appears.  
With a short bond dissipation quench, however,  
the relaxation of $\rho_1$ is accelerated, so that the initially more localized state relaxes faster, signaling the QME, as shown by the blue dashed line in Fig. \ref{fig2}(b). Figure \ref{fig2}(c) 
confirms that this effect originates from the suppression of the slowest-mode 
amplitude $|\mu_1|$ during the quench. We note that in Fig. \ref{fig2}, we fix $a=1$ throughout. Choosing $a=-1$ leads to qualitatively similar results.

\begin{figure}[tb]
	\centering
    \includegraphics[width=8.57cm]{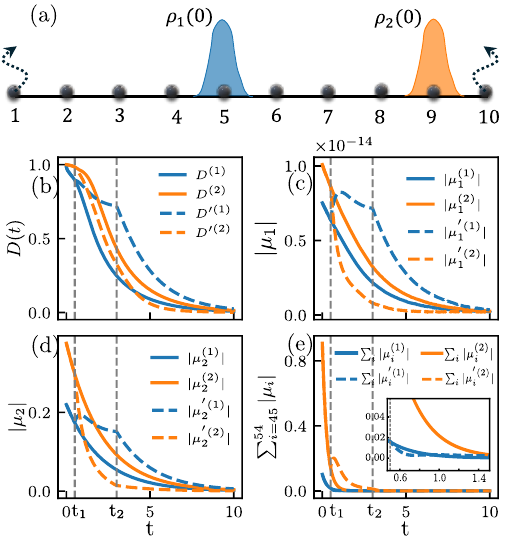}
    \caption{Boundary dissipation with a temporary bond-dissipation quench.  
    	(a) Initial states $\rho_1(0)=|5\rangle\langle 5|$ and $\rho_2(0)=|9\rangle\langle 9|$.  
    	(b) Time evolution of the trace distance $D^{(i)}(t)$. 
    	With bond dissipation, $\rho_2$ relaxes faster for $a=-1,p=2$ (orange dashed line, QME), 
    	while $\rho_1$ relaxes more slowly for $a=1,p=2$ (blue dashed line, anti-QME).  
    	(c) Time evolution of the slowest-mode amplitude $|\mu_1|$ for the cases in (b).  
    	(d) Time evolution of the next-slowest mode amplitude $|\mu_2|$.  
    	(e) Time evolution of intermediate modes \(|\mu_j|\) with \(j = 45\) to \(54\). 
    	 Here $\Gamma=0.4$,  $\gamma^b_1=\gamma^b_L=0.2$, $t_1=0.5$ and $t_2=3$.}
	\label{fig3}
\end{figure}

\textit{Example II: Boundary loss.—} As a second example, we consider intrinsic particle loss acting only on the two edge sites~\cite{PTopen1,boundary1,boundary2,boundary3}, described by
\begin{equation}
	O^{(0)}_1 = \sqrt{\gamma^b_1} c_1,
	\quad
	O^{(0)}_L = \sqrt{\gamma^b_L} c_L,
\end{equation}
with dissipation strengths $\gamma^b_1$ and $\gamma^b_L$.  We set the system size to $L=10$ and compare two initial states [Fig. \ref{fig3}(a)]: a single-particle state localized at site $5$ ($\rho_1$), and another localized at site $9$ ($\rho_2$).
Without bond dissipation, Fig. \ref{fig3}(b) shows that $\rho_1$, being closer to 
the boundary, relaxes faster than $\rho_2$, and no QME appears.  
When bond dissipation $O^{(1)}_j$ is switched on within a  finite time window
$t_1 < t < t_2$, the relaxation hierarchy can be reversed.
As shown in Fig. \ref{fig3}(b), when starting from $\rho_2$, applying bond dissipation with parameters
$a=-1$, $p=2$ accelerates the relaxation, giving rise to the QME (orange dashed line).
In contrast, when starting from $\rho_1$, bond dissipation with $a=1$, $p=2$ slows down the relaxation, leading to an anti-QME (blue dashed line).

The underlying mechanism is revealed by examining Liouvillian modes. 
Figure \ref{fig3}(c) shows the slowest-mode coefficient  $\mu_1(t)$. Although its evolution reflects both QME and anti-QME, its magnitude is vanishingly small ($\sim 10^{-14}$), indicating negligible overlap with the initial states. The relaxation is instead dominated by the next-slowest mode $\mu_2(t)=\mathrm{Tr}[\, l_2^\dagger \rho(t) \,]$, as shown in Fig. \ref{fig3}(d).
The bond dissipation selectively suppresses (for $a=-1$) or enhances (for $a=1$) the amplitude of this dominant slow mode, directly determining the emergence of QME or anti-QME.
Finally, Fig. \ref{fig3}(e) illustrates the contributions from several intermediate modes.
Although their coefficients evolve with trends opposite to that of $\mu_2$,
implying that weight reduced in the slow mode is redistributed into these modes,
their larger decay rates render their contributions negligible for the overall relaxation time.

%to redistribute the weight of slow modes and thereby control whether the system exhibits the QME or the anti-QME. 

The effect induced by the above bond dissipation originates from the fact that the slowest mode of the boundary-loss system (whose weight is negligibly small) and the next-slowest mode both exhibit in-phase amplitudes between next-nearest-neighbor sites. To gain an intuitive understanding of this mechanism, and without loss of generality, we consider systems governed by the simplest tight-binding Hamiltonian,
\begin{equation}\label{ham1}
	H = \sum_{j=1} J \left( c_j^\dagger c_{j+1} + c_{j+1}^\dagger c_{j} \right),
\end{equation}
where $J$ is the hopping amplitude, set to unity throughout this work. Before entering the explicit examples, we first clarify how the bond dissipation reshapes the relaxation dynamics of an open system, and outline the underlying mechanism from the viewpoint of eigenstate phase structure. For illustrative purposes, we temporarily impose periodic boundary conditions. The single-particle eigenstates of Eq.~(\ref{ham1}) are plane waves $|k\rangle = \tfrac{1}{\sqrt{L}} \sum_{j=1}^L e^{ikj} |j\rangle$, with eigenvalues $E_k = 2J \cos k$, where the allowed momenta are $k = 2\pi n/L$ with integer $n \in (-L/2, L/2]$. It is straightforward to see that the phase difference between two sites separated by a distance 
$p$ is $pk$, and therefore varies across the band. At the top of the band ($k=0$), all lattice sites are in phase; at the bottom of the band ($k=\pi$), neighboring sites are out of phase while next-nearest neighbors are in phase. For intermediate momenta, e.g., at $k=\pi/2$, the next-nearest neighbors instead become out of phase. Bond dissipation couples sites separated by a given distance ($p=1$ for nearest neighbors and $p=2$ for next-nearest neighbors), and thereby favors specific relative phase configurations. Modes whose phase structure conflicts with the dissipative constraint are selectively suppressed, while phase-compatible modes can be enhanced. This mechanism provides a direct and controllable means to suppress or amplify slow relaxation channels, enabling either the QME or the anti-QME depending on the dissipative protocol.

\begin{figure}[tb]
	\centering
	\includegraphics[width=8.57cm]{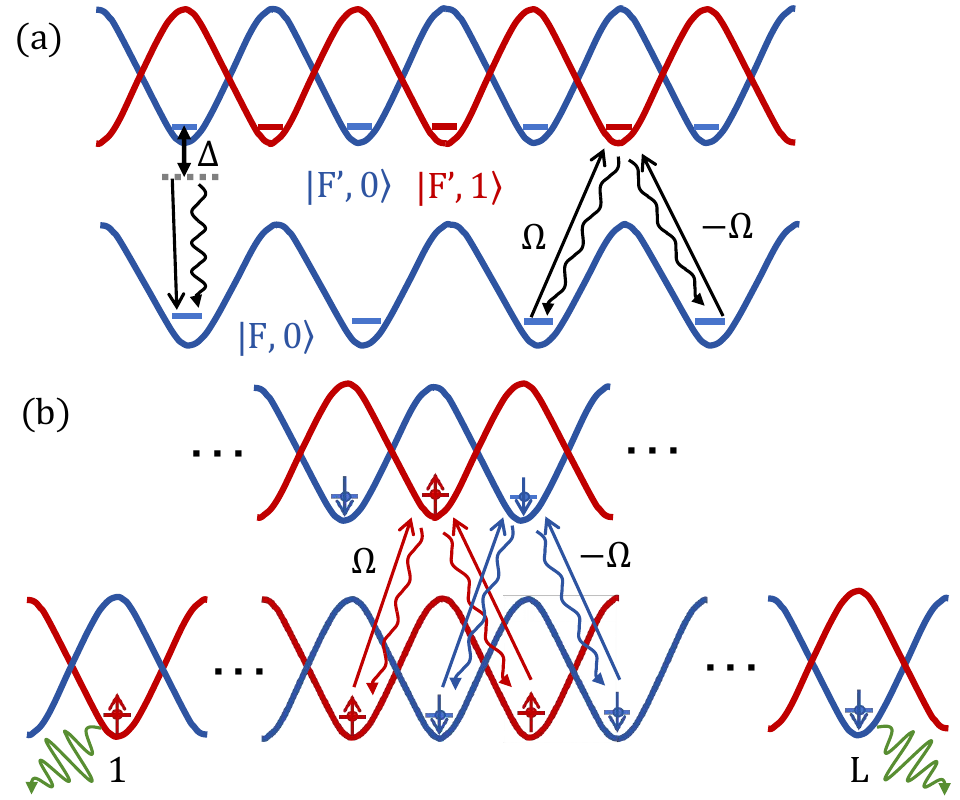}
	\caption{Experimental schemes.
		(a) Realization of local dephasing (via a weak $\pi$-polarized beam) and nearest-neighbor ($p=1$) bond dissipation (via $\sigma^+$-polarized driving), using a state-dependent auxiliary lattice.
		(b) Setup for boundary loss combined with next-nearest-neighbor ($p=2$) bond dissipation, employing a spin-dependent ground-state lattice and an auxiliary lattice shifted by half a period.	}
	\label{fig4}
\end{figure}

\textit{Experimental realization.—} We now discuss feasible schemes to implement the two dissipative mechanisms introduced above in an optical lattice loaded with ultracold atoms. The relevant atomic levels are characterized by the total hyperfine angular momentum $F$ and its magnetic sublevel $m_F$. To realize independent dissipative channels, bond dissipation with $p=1$ is implemented by driving $\ket{F,m_F=0}\to\ket{F',m_F=+1}$ using $\sigma^+$-polarized light, while local dephasing is induced by a weak $\pi$-polarized beam driving $\ket{F,m_F=0}\to\ket{F',m_F=0}$, and state-dependent optical lattices ensure aligned ground and excited states in the dephasing channel while shifting the excited-state lattice by half a period in the bond channel, as shown in Fig. \ref{fig4}(a).
For bond dissipation, the driving wavelength is set to twice the lattice constant, so that neighboring sites acquire opposite Rabi phases, $+\Omega$ and $-\Omega$.  
This spatially antisymmetric coupling generates the annihilation part of the bond jump operator $(c_j-c_{j+1})$, whereas the subsequent collective spontaneous emission from $\ket{F',m_F=+1}$ in the long-wavelength regime produces an approximately symmetric creation channel proportional to  $(c^{\dagger}_j+c^{\dagger}_{j+1})$~\cite{PTopen13,PZollerB0,PZollerB1,PZollerB2,PZollerB3,PZollerB4,SM}.  The effective dissipation rate scales as $\Gamma \sim \Omega^2/\Gamma_e$~\cite{SM}, tunable via the Rabi frequency $\Omega$ and excited-state spontaneous emission rate $\Gamma_e$, and can be switched on or off simply by controlling the driving beam. According to the conditions for adiabatic elimination, we require $\Gamma_e \gg \Omega, J$, where \(J\) is the ground-state hopping amplitude. Under these conditions, the excited-state population is suppressed to \(\sim \Omega^2/\Gamma_e^2 \ll 1\), and the ground-state dynamics is slow compared to the excited-state decay.  More details are presented in Supplementary Material~\cite{SM}.
%The effective dissipation rate can be tuned via the Rabi frequency $\Omega$ and the excited-state linewidth $\Gamma_e$, scaling as $\Gamma \sim \Omega^2/\Gamma_e$.  In practice, bond dissipation can be switched on and off simply by applying or removing the corresponding driving light.
Dephasing dissipation is realized using a far-detuned beam with detuning $\Delta$ ($|\Delta|\!\gg\!\Gamma_e$), so that atoms are only virtually excited before decaying back to the ground manifold~\cite{Uys2010PRL,Kuhr2005PRA,dephasingex}.  
The associated random photon recoils introduce stochastic phase shifts, producing pure dephasing without affecting on-site populations.  
The corresponding dephasing rate is given by
$\gamma \approx \frac{\Gamma_e}{2}\,\frac{s_0}{1+s_0+(2\Delta/\Gamma_e)^2}$,
where $s_0$ is the on-resonance saturation parameter set by the imaging beam intensity. To prevent crosstalk between the two dissipative channels, each driving field is far detuned from the other channel's transition: the $\sigma^+$
field is off-resonant for the $\pi$ transition, and the $\pi$ field is off-resonant for the $\sigma^+$
transition. This large detuning, combined with the distinct polarizations and atomic selection rules, ensures that each drive only affects its intended channel.

Since spontaneous emission may redistribute population across Zeeman sublevels, continuous Raman sideband cooling (RSC)~\cite{RSC1,RSC2} is applied to rapidly repump atoms back into the target state $\ket{F,m_F=0}$, thereby ensuring that only the $m_F=0$ ground state participates in the dissipative dynamics. RSC repumps atoms back to \(|F,m_F=0\rangle\) at a rate satisfying $\gamma_{\rm leak}\lesssim\gamma_{\rm RSC}\ll\Gamma,\gamma^d$, where \(\gamma_{\rm leak}\) denotes the effective leakage rate from the target subspace. Under this condition, RSC stabilizes the target manifold without competing with the engineered dissipative dynamics.

\begin{figure}[tb]
	\centering
	\includegraphics[width=8.57cm]{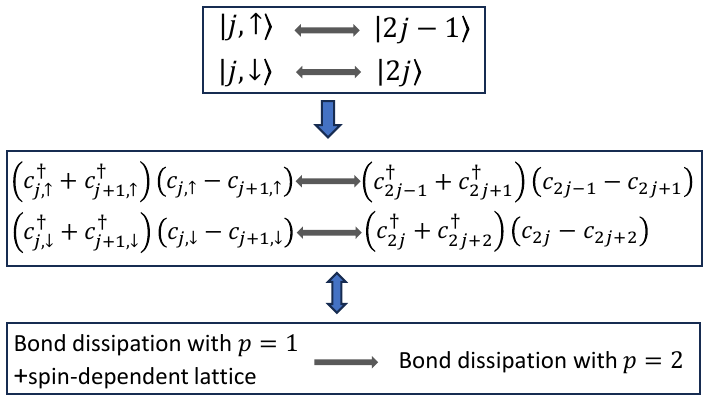}
	\caption{Schematic illustration of the route to realizing bond dissipation with $p=2$. By preparing a spin-dependent lattice, spin-up and spin-down states are mapped onto odd and even lattice sites, respectively. The spin-dependent 
	$p=1$ bond dissipation then effectively implements $p=2$ bond dissipation. }
	\label{fig5}
\end{figure}
For boundary loss combined with $p=2$ bond dissipation [Fig. \ref{fig4}(b)], 
a spin-1/2 lattice is encoded in two hyperfine states $|\!\uparrow\rangle=|F_1,m_{F1}\rangle$ and $|\!\downarrow\rangle=|F_2,m_{F2}\rangle$, with odd (even) sites mapped to spin-up (spin-down). 
A Raman coupling between a standing wave and a plane wave generates a deep spin-dependent ground-state lattice $V_p(x)\sigma_z$, where spin-conserving tunneling is inhibited and spin-flip processes realize the hopping term of Hamiltonian~(\ref{ham1})~\cite{PTopen13,WangYC}.  
To implement $p=2$ bond dissipation, an auxiliary spin-dependent lattice shifted by half a period is introduced,
constructed from two hyperfine states $|\!\uparrow\rangle=|F'_1,m'_{F1}\rangle$ and $|\!\downarrow\rangle=|F'_2,m'_{F2}\rangle$ chosen to satisfy $m_{F1}-m'_{F1}=m_{F2}-m'_{F2}$.  
The driving laser polarization is chosen to achieve state-selective coupling between the ground and auxiliary lattices, with its wavelength set to twice that of the standing-wave laser to introduce the necessary $\pi$ phase shift in the effective Rabi frequency. Then, the creation part of bond dissipation with 
$p=2$ is also generated by the isotropy of spontaneous emission. As illustrated in Fig. \ref{fig5}, the principle for generating bond dissipation with $p=1$ is combined with a spin-dependent lattice to produce bond dissipation with $p=2$. 
Boundary-localized particle loss can be engineered through well-developed approaches, such as employing tightly focused electron beams~\cite{Weitenberg2011Nature,Bakr2009Nature,Ott2008,Ott2013} or femtosecond laser pulses~\cite{Wessels2018} to ionize atoms; driving photoassociation processes that convert atoms into molecules~\cite{Tomita2017} or inducing decay into molecular channels~\cite{Syassen2008,Amico2018}; using near-resonant light scattering to impart sufficient recoil energy for atoms to escape from the trap~\cite{Pfau1994,Patil2015,Bouganne2020,Corman2019,Lebrat2019,Huang2023}; or via photon-scattering–induced band excitations that populate weakly confined higher bands and cause loss~\cite{dephasingex}.

\textit{Conclusion.—}
We have established a general strategy to realize QME and anti-QME in open quantum systems by harnessing bond dissipation. A temporary bond-dissipation quench redistributes spectral weight among Liouvillian modes, enabling controllable suppression or enhancement of slow relaxation channels. Importantly, this mechanism is independent of specifically chosen systems and initial states, thereby overcoming a fundamental limitation of earlier QME scenarios.
We verified the universality of this approach in systems with dephasing and boundary dissipation, and outlined realistic schemes for cold-atom implementation. Our findings not only deepen the understanding of nonequilibrium relaxation in open quantum systems, but also establish bond dissipation as a versatile tool for dynamical control, with promising applications in accelerated state preparation and dissipative quantum technologies.

\begin{acknowledgments}
	This work is supported by National Key R\&D Program of China under Grant No.2022YFA1405800, the Key-Area Research and Development Program of Guangdong Province (Grant No.2018B030326001), Guangdong Provincial Key Laboratory(Grant No.2019B121203002). Y. Liu is also supported by the Postdoctoral Fellowship Program of CPSF under Grant Number GZB20250776.
\end{acknowledgments}
%\bibliographystyle{apsrev4-2}
%\bibliography{references}  % 不要加 .bib 后缀

\end{document}